\documentclass{elbioimp2}
\usepackage[utf8]{inputenc}

\usepackage[backend=biber,style=vancouver]{biblatex}
\usepackage{csquotes}
\usepackage{amsmath}
\usepackage{xcolor}
\title{Machine Learning based Pointing Models for Radio/Sub-millimeter Telescopes}
\shorttitle{ML-based pointing correction for radio and sub-mm telescopes}
\author{Bendik Nyheim\affiliation{Department of Physics, University of Oslo, N-0316 Oslo, Norway} \affiliation{Institute of Theoretical Astrophysics, University of Oslo, P.O. Box 1029, Blindern 0315, Oslo, Norway}, Signe Riemer-Sørensen\affiliation{Analytics and AI, SINTEF Digital, P.O. Box 124 Blindern, N-0314 Oslo, Norway; E-mail any correspondence to: Signe.Riemer-Sorensen@sintef.no},  Rodrigo Parra\affiliation{European Southern Observatory, Alonso de Cordova 3107, Vitacura, Santiago, Chile}, and Claudia Cicone\affiliation{Institute of Theoretical Astrophysics, University of Oslo, P.O. Box 1029, Blindern 0315, Oslo, Norway}} 
\shortauthor{Nyheim et al.}
\elbioimpreceived{12 02 2024}  
\elbioimppublished{Xx XX 2024}
\elbioimpfirstpage{1}
\elbioimpvolume{10}
\elbioimpyear{2024}
 
\begin{document}
\maketitle

\begin{abstract}
Radio, sub-millimeter and millimeter ground-based telescopes are powerful instruments for studying the gas and dust-rich regions of the Universe that are invisible at optical wavelengths, but the pointing accuracy is crucial for obtaining high-quality data. Pointing errors are small deviations of the telescope’s orientation from its desired direction. The telescopes use linear regression pointing models to correct for these errors, taking into account various factors such as weather conditions, telescope mechanical structure, and the target’s position in the sky. However, residual pointing errors can still occur due to factors that are hard to model accurately, such as thermal and gravitational deformation and environmental conditions like humidity and wind. Here we present a proof-of-concept for reducing pointing error for the Atacama Pathfinder EXperiment (APEX) telescope in the high-altitude Atacama Desert in Chile based on machine learning. Using historic pointing data from $2022$, we trained eXtreme Gradient Boosting (XGBoost) models that reduced the Root Mean Squared Error (RMSE) for azimuth and elevation (horizontal and vertical angle) pointing corrections by $4.3\%$ and $9.5\%$, respectively, on hold-out test data\footnote{The code is available here \url{https://github.com/benyhh/ml-telescope-pointing-correction}}.
Our results will inform operations of current and future facilities such as the next-generation Atacama Large Aperture Submillimeter Telescope (AtLAST).
\end{abstract}

\section{Introduction}
Radio, millimeter and sub-millimeter (radio/(sub)-mm) telescopes are designed to collect and detect electromagnetic radiation at wavelengths longer than $\sim300~\mu$m, up to $\lambda \sim 15$~m, corresponding to frequencies from $\nu\sim 1$~THz to a few MHz \cite{Corstanje2017}. The short wavelength cutoff, corresponding to the division between far-infrared (FIR) and sub-mm regime where the former can only be observed from space observatories and the latter only from very high and dry places on Earth \cite{app132111706, AandA2006}, is determined by the high content in water vapor (H$_2$O) and other molecules of the Earth's troposphere which absorb infrared and sub-mm photons. The long wavelength cutoff is due to the interaction between low frequency radio waves and charged particles in the ionosphere \cite{app132111706}, which becomes non negligible at frequencies below 10 MHz. 

Through radio and sub-mm observations astronomers can peer into the most dust-obscured and gas-rich regions of galaxies which are invisible at other wavelengths, hence learning how stars (and their surrounding planetary systems) form out of cold and dense molecular clouds \cite{Omont07, Tacconi+20}, and studying the chemical complexity of interstellar gas and planetary atmospheres through sub-mm molecular line observations \cite{Jorgensen+20}. Due to the expansion of the Universe, the wavelength of the radiation emitted from the first galaxies that formed after the Big Bang is shifted towards longer wavelengths, hence their rest-frame infrared emission from dust and cold gas can be captured only through sub-mm observations \cite{Carilli+Walter13}. The first image of a super-massive black hole's event horizon, and of its gravitationally lensed photon capture ring, was obtained through observations conducted at $\lambda = 1.3~mm$ with a network of (sub)-mm ground-based telescopes which were all simultaneously pointing towards the center of the galaxy M87, and whose data were correlated through the Very Large Baseline Interferometry technique \cite{EHT2019}. These are only a few examples of scientific drivers of sub-mm and radio astronomical observations, for a more comprehensive overview we refer the reader to the white papers of current and future sub-mm and radio observatories \cite{Mroczkowski+23, Ramasawmy+22, Klaassen+20, ALMAbook2023, MurphyngVLA17, FerrariSKA}.

\begin{figure}[htp]
   \centering
   \includegraphics[width=\columnwidth]{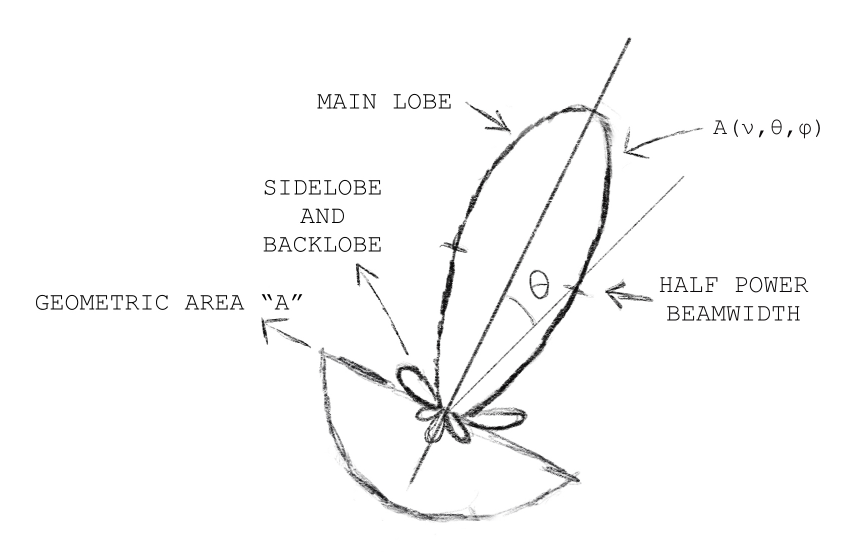}
   \caption{Sketch of the beam of a telescope.}
   \label{fig:beamwidth}
\end{figure}

One of the key components of a radio/ (sub)-mm antenna is its reflective surface, which collects and focuses the incoming radiation.
Most radio/(sub)-mm telescopes have a large, parabolic primary mirror (also known as primary dish), reflecting incoming radiation onto a smaller, secondary mirror (also known as subreflector), and often additional optical elements (tertiary mirror, etc.) to improve imaging capabilities and/or direct the beam of radiation onto different instrument cabins.
After being reflected by the last optical element, e.g. the subreflector in the case of a two-mirror telescope, the radiation reaches a detector or receiver placed at the focal point of the telescope, which records and processes the signal.
Throughout this paper, we will use the term `pointing' to refer to the process of positioning the telescope to observe a portion of the celestial sphere, i.e. the astrophysical target of interest.
The goal of pointing is to overlap the center of the astrophysical source with the center of the resolution element or `beam' of the telescope, 
where the antenna power pattern peaks and so the sensitivity of the telescope is maximal (see Figure \ref{fig:beamwidth}).
The angular size of the beam, i.e. the (main) beam size, can be quantified by its half power beam width  (HPBW):
\begin{equation}\label{eq:beam_size}
    {\rm HPWB}_{mb} = \theta_{mb} \sim k\frac{\lambda}{D},
\end{equation}
where $\lambda$ is the observed wavelength, $D$ is the diameter of the primary mirror, and $k$ is a numerical factor that accounts for the non-uniform illumination of the main dish (e.g. blockage due to the subreflector shade, taper, etc.). Equation~\ref{eq:beam_size} shows that the beam becomes smaller when observing shorter wavelengths and when using larger telescopes, hence a higher pointing accuracy is needed in these situations. 

Radio/(sub)-mm telescopes integrate the signal over time to create a composite image or spectrum.
This process requires highly accurate pointing, as even slight errors in the telescope's orientation can significantly affect the resulting data quality.
Pointing errors, often referred to as pointing offsets, can be caused by various factors, including thermal deformation of the telescope components \cite{Dong2018},
gravitational deformation \cite{GravDeformation}, and other environmental factors like humidity \cite{Corstanje2017} and wind \cite{Gawronski2005}.
Most current radio/(sub)-mm instruments are single-pixel and so do not have imaging capabilities:
the receiver images only the central resolution element of the telescope and not the full field of view (FoV). In this case, any radiation received from a portion of the sky that lies outside such beam would not be processed by the instrument. For this reason, reducing the pointing error is crucial for single-pixel receivers. Some telescopes have an optical pointing camera, however most of the astrophysical targets of sub-mm and radio observations are very faint or invisible at optical wavelengths.
Hence, the correct positioning of the source within the resolution element (beam) and at the center of the field of view cannot be checked directly during observation.
To achieve sufficient accuracy, radio/(sub)-mm telescopes use pointing models \cite{stumpff1972}, which take into account a range of factors that can contribute to the pointing error,
including weather conditions, telescope structure, and the target's position in the sky.

The Atacama Pathfinder EXperiment (APEX) telescope \cite{APEX2006}\footnote[1]{\href{http://www.apex-telescope.org/ns/}{apex-telescope.org/ns/}} located at an altitude of 5100m on the Chajnantor Plateau in the Chilean Atacama Desert, is currently the largest single-dish sub-millimeter telescope in the southern hemisphere. APEX hosts multiple instruments observing from $\sim180$~GHz up to $\sim850$~GHz. Although APEX's maximum field of view is larger than 10~arcmin, most of the instruments with spectroscopic capabilities (heterodyne receivers) are single-beam.  The APEX 12-meter diameter translates into a beam size ranging from about 40~arcsec to 7~arcsec, depending on the observed frequency.

APEX currently uses an effective analytical pointing model that is recalibrated periodically through pointing measurement campaigns at regular intervals (3-6 months).
However, observations of known point sources reveal a remaining residual pointing offset (with a median value of $2.84$ arcsec and an interquartile range of $1.55$ to $4.66$ arcsec at $230$~GHz) whose origin is not understood but attributed to oversimplification of the pointing model and non-linear mechanical/thermal effects that are difficult to model.

This paper aims to explore the use of machine learning to increase the accuracy of the current pointing model at APEX, based on observational data such as weather patterns and telescope pointing.
A machine learning approach would in particular benefit larger radio/(sub)-mm telescopes such as the Atacama Large Aperture Submillimeter Telescope (AtLAST)\footnote[2]{Link to AtLAST website \href{https://www.atlast.uio.no/}{atlast.uio.no}}.
AtLAST will have a $50$-meter diameter primary mirror and $12$-meter diameter subreflector. Hence, the subreflector will be as big as APEX's primary mirror.
Due to the size and complexity of AtLAST, it will be harder to develop an analytical pointing model.
By developing a more advanced and reliable pointing model with machine learning,
this research can enhance the capabilities of current and future radio/(sub)-mm telescopes.

\section{The pointing model}\label{sec:pt_model}
Even the brightest sub-mm/radio astrophysical sources in the Universe emit a signal that is much weaker than the emission from the Earth's atmosphere,
which means that the signal is hidden in the noise and needs to be extracted using long integrations and modulation techniques.
Therefore, astronomers must know that the pointing is accurate before initiating a long integration on the source.

The pointing model at APEX consists of two steps:  (i) an analytical model calibrated to specific sources, and (ii) additional pointing corrections performed at regular intervals based on recently observed pointing offsets (difference between input coordinates and the observed coordinates of the source). This pointing model is implemented as part of the APEX control system (APECS) \cite{APECS2006}.
The analytical model consists of fitting multiple terms to the many measurements of the pointing offset.
These terms include geometric terms motivated by the mechanical structure or terms related to, for example, temperature measurements.
The fitted terms are used for $1$-$2$ months, while the model runs in the background adjusting all input coordinates.
The additional pointing corrections (step (ii)) are performed by the observers every $1$-$2$ hours during science observations and/or before observing a new target.

During operations, astronomical observatories adopt the local Altitude-Azimuth (AltAz) coordinate system to indicate the position of a source on the sky. The sky can be represented as the internal surface of an imaginary sphere that encloses the Earth, the so-called Celestial sphere. On this surface, one can draw the two perpendicular great circles that are used as a reference in the AltAz system: the local horizon and the local meridian. The position of each point on the celestial sphere can then be determined through two angular coordinates: the Azimuth (Az) and the Elevation (El). The azimuth is the horizontal coordinate, ranging from 0$^{\circ}$ to 360$^{\circ}$, usually defined to increase from South to North moving towards East. The Elevation (or altitude) is the vertical coordinate that measures the angular distance of a point from the horizon, and it varies from 0$^{\circ}$ to 90$^{\circ}$, where a source at elevation 0$^{\circ}$ is on the horizon, and at 90$^{\circ}$ is at Zenith (vertical above the observer).  

The APEX pointing correction procedure can be described as:
\begin{equation}
    Az = Az_{input} + \Delta Az_{analytical model} + \Delta Az_{correction} 
\end{equation}
\begin{equation}
    El = El_{input} + \Delta El_{analytical model} + \Delta El_{correction}.
\end{equation}
Here, $Az_{input}$ and $El_{input}$, indicate the input coordinates of the target.
The second terms $\Delta Az_{analytical model}$ and $\Delta El_{analytical model}$ are the positions adjustments determined by the analytical model, while $\Delta Az_{correction}$ and $\Delta El_{correction}$, are the corrections based on a recently measured pointing offset.

\subsection{Pointing Corrections} 
Although the analytical linear regression pointing model at APEX fits pointing campaign observations well,
it cannot accurately model the changes in pointing over time, resulting in residual errors.
APEX observers regularly (every $1$-$2$ hours) correct these errors by observing a strong source with known coordinates,
measuring the pointing offset, and updating the model's terms for azimuth and elevation correction ($\Delta Az_{correction}$ and $\Delta El_{correction}$, respectively).
This procedure is referred to as a pointing scan, and the pointing model is updated as follows:

\begin{equation}
    \Delta Az_{correction} = \Delta Az_{correction} + \delta_{Az} \label{eq:ca},
\end{equation}
\begin{equation}
     \Delta El_{correction} = \Delta El_{correction} - \delta_{El},\label{eq:ie}
\end{equation}
where $\delta_{Az}$ and $\delta_{El}$ are the recently observed pointing offsets in azimuth and elevation, respectively.
The observers perform these pointing corrections every couple of hours to ensure the pointing accuracy is sufficient during science observations.

\section{Data and Features}
This work makes use of historical APEX data from 2022 measured by various sensors mounted on the telescope structure, in addition to information about the continuous pointing corrections applied by various telescope systems. The data are collected and stored locally at the telescope and are not publicly available This is a list of the data employed in our analysis:
\begin{itemize}
    \item Weather data such as wind speed, wind direction. humidity, pressure, dewpoint from the APEX weather station. Sampled at $5$ data points per minute.
    \item Temperature sensors outside and inside the telescope structure at various locations. Sampled at $2$ to $12$ data points per minute depending on location.
    \item Position and rotation of the secondary mirror. Sampled at 6 data points per minute.
    \item Displacement measurements of the yoke. Sampled at 12 data points per minute.
    \item Tiltmeters mounted on the telescope. Sampled at 12 data points per minute.
    \item Automatic adjustments made based on temperatures, tiltmeters, displacement of yoke, measurements of quadrupod (structure supporting subreflector) sag. Sampled at 12 data points per minute.
    \item Input coordinates. Sampled at 6 data points per minute.
    \item Coordinates after applying the underlying pointing model. Same sample times as the coordinates.
    \item Time based features.
    \item Pointing offsets observed during pointing scans.
    \item Position of the sun.
\end{itemize}

\subsection{Instruments}
The APEX instruments operate at different frequencies. In Table \ref{tab:instrument_usage} we list the instruments whose data have been used in our work, together with their frequency coverage and,
the number of available pointing scans. These instruments were chosen based on the number of pointing scans available, and the reliability of the data. A `scan' is the shortest possible observation recorded by the telescope. Scans of bright sources are frequently taken to calibrate the telescope, and each scan provide one pointing error data point for both the azimuth and elevation angle.
We perform two separate analyses; one using only pointing scans from NFLASH230 and one including all listed instruments.

\begin{table}[htp]
    \centering
    \begin{tabular}{lcr}
        Instrument & Frequency band [GHz] &\# of scans \\
        \hline
        NFLASH230 & $200$-$270$ &3197 \\
        LASMA345  & $268$-$375$ &1861 \\
        NFLASH460 & $385$-$500$ &1394 \\
        SEPIA660  & $578$-$738$ & 856 \\
        SEPIA345  & $272$-$376$ & 818 \\
        SEPIA180  & $159$-$211$ & 359 \\
        \hline
    \end{tabular}
    \caption[Number of scans for each instrument]{The instruments, their frequency bands\footnote{Detailed descriptions of each instrument available at \href{https://www.eso.org/sci/facilities/apex/cfp/cfp110/instrument_summary.html.html}{website}}, and number of pointing scans performed in $2022$ with each instrument. There are $8485$ scans in total.}
    \label{tab:instrument_usage}
\end{table}
\subsection{Data Processing}
Using clean data is essential when training a machine learning model. The dataset was cleaned based on several conditions, informed by domain knowledge from APEX staff. The following criteria were used to exclude unsuitable scans:
\begin{itemize}
\item Scans from January and February 2022, as weather during this period is unreliable and the number of scans is limited.
\item Scans after September 17, 2022, as we do not have sensor data beyond this date to correlate with the pointing scans data.
\item Scans that are tracking tests or those using the Moon as the source do not accurately represent the pointing offsets.
\item Scans with a high uncertainty for the measurement of $\Delta Az_{correction}$ and $\Delta El_{correction}$.
\end{itemize}
After this filtering, there were 5901 out of 8485 scans left covering the period from March to September 2022.

\subsection{Data splits}
\begin{figure*}[htp]
    \centering
    \includegraphics[width=\textwidth]{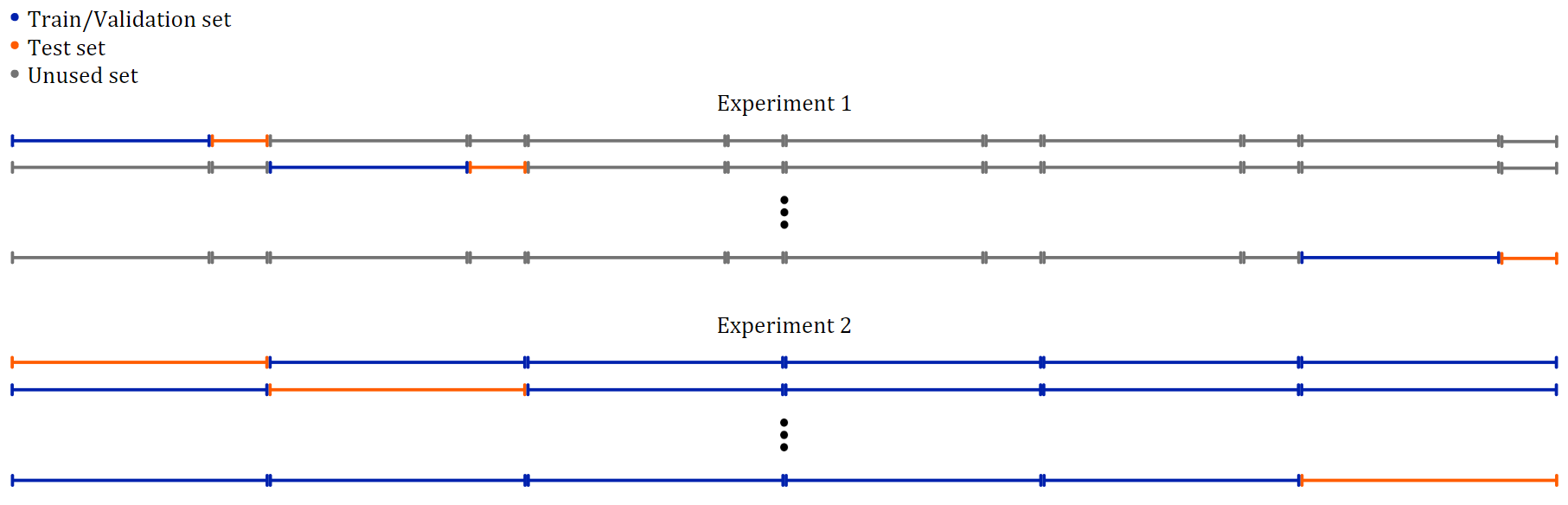}
    \caption[Data split experiments for pointing correction model]{Data splits for the pointing correction model. In \textbf{Experiment 1}, the data is split into six equal-sized folds sorted by date.
    For each fold, we use the last part (orange) for testing and the remaining part (blue) for training and validation, while the rest (grey) is unused for that fold.
    In \textbf{Experiment 2}, the data is split into six equal-sized folds sorted by date. However, we use one whole fold for testing this time (organge) and the remaining five for training and validation (blue). This process is repeated six times, with each fold used exactly once for testing.}
    \label{fig:datasplit_cases}
\end{figure*}

Our goal is to train a model to predict the offset, $\delta_{Az}$ and $\delta_{El}$, that would be observed during a pointing scan under some given conditions.
We hypothesise that the pointing error is being affected by factors that are independent of time/long-term history such as weather, and factors that may change (slightly) over time for example the mechanical structure. Hence, we investigate whether models trained on smaller, more recent datasets outweigh the advantage of training on all data. This leads to the use of different data splits:
\begin{itemize}
    \item \textbf{Experiment 1:} The dataset is sorted by date and split into six equal-sized folds. We consider each of the folds one by one. For each of these folds, we use the last $1/6$th of the data as a test set and the remaining $5/6$th for training and validation.
    \item \textbf{Experiment 2:} The dataset is sorted by date and split into six equal-sized folds. We use $5/6$ of the data for training and validation and the remaining for testing.
    We repeat this process six times, using each fold for testing once.
\end{itemize}
Figure \ref{fig:datasplit_cases} illustrates the two experiments.
In both experiments, we performed cross-validation with $5/6$ of the data for training and validation, and tested on the last $1/6$.

We partitioned the data so that pointing scans from a given day are exclusively included in either the training set or the validation set, and not both.
During the data split, we used $67\%$ for training and $33\%$ for validation.

\subsection{Model Architecture}
Due to its overall good performance on tabular data \cite{2021arXiv210603253S}, we utilized the decision tree-based method XGBoost \cite{Chen2016} for modelling, with the Root Mean Squared Error (RMSE) as loss metric \footnote{The code and hyperparameter optimization is available here \url{https://github.com/benyhh/ml-telescope-pointing-correction}}. 
For each model (for each cross-validation experiment, target variable, and set of features), the hyper parameters were optimized individually through a Bayesian search of 200 different combinations using HyperOpt \cite{Bergstra2015hyperopt}. The search space is given in Table \ref{tab:xgb_hyperparameters_pcorr}.

\begin{table}[htp]
    \centering
    \begin{tabular}{lcc}
        Parameter & Sample Distrib. & Range \\ \hline
        \texttt{max\_depth} & Uniform & [$1$, $5$] \\ 
        \texttt{colsample\_bytree} & Uniform & [$0.5$, $1$] \\ 
        \texttt{n\_estimators} & Uniform & [$20$, $500$] \\ 
        \texttt{learning\_rate} & Log-Uniform & [$10^{-5}$, $1$] \\ 
        \texttt{subsample} & Uniform & [$0.5$, $1$] \\ 
        \texttt{min\_child\_weight} & Uniform & [$1$, $10$] \\
        \hline
    \end{tabular}
    \caption[Hyperparameter search space for XGBoost pointing correction model]{List of hyper parameters sampled during tuning for the XGBoost pointing correction model. The table includes names, initial distributions, and corresponding ranges.}
    \label{tab:xgb_hyperparameters_pcorr}
\end{table}

\subsection{Feature Selection}
XGBoost can only learn one target variable, so azimuth and elevation were considered independently.
Given the large number of possible variables, we risk introducing noise by including less relevant features. To reduce the noise, we investigated multiple sets of input features $k=[2,5,10,20,30,40,50]$, selecting the $k$ features with the greatest mutual information (MI) \cite{Shannon1948} with the target variable. The MI was computed on the combined training and validation data for a given fold, before fitting the XGBoost models. In addition to the MI feature selection, we added the input coordinates $Az_{input}$ and $El_{input}$ as features for all models.

\subsection{Model Evaluation}
We evaluated a model's performance by comparing the RMSE on each test fold with the observed RMSE of the existing pointing model.
The metric we used for this is the RMSE ratio between the XGBoost pointing correction model and the current model, given by
\begin{equation}\label{eq:rms_compared}
    r_{RMS,j} = \frac{RMS_{target,j}}{RMS_{current,j}}.
\end{equation}
This measure is useful because it compares the model's performance to the current performance of the telescope.
If $r_{RMS,j}<1$, it indicates that the XGBoost model provides an improvement over the current performance of the telescope for a given fold.

To obtain an overall measure of the performance of the XGBoost approach, we averaged the ratios $r_{RMS,j}$ over all six test folds
\begin{equation} \label{eq:mean_rms_compared}
    \bar{r}_{RMS} = \sum_{i=1}^6 \frac{RMS_{model,j}}{RMS_{current,j}}.
\end{equation}
This gives us an average ratio $\bar{r}_{RMS}$, which measures the expected performance of the pointing correction model.
If $\bar{r}_{RMS} < 1$, it indicates that the XGBoost models outperforms the current pointing correction method on average across all test folds.
By comparing the average ratio $\bar{r}_{RMS}$ for the two different cross-validation experiments in Figure \ref{fig:datasplit_cases},
we can identify which data-splitting approach leads to better generalization.

\section{Results}

Tables \ref{tab:exp1_minval_nflash230} and \ref{tab:exp2_minval_nflash230} presents the validation and test RMSE ratios for all folds of the NFLASH230 model in Experiment 1 and 2. In Experiment 1, the NFLASH230 model shows promising performance on validation data with mean RMSE ratios of $0.858$ ($\sigma=0.064$) for azimuth and $0.810$ ($\sigma=0.040$) for elevation. However, this did not generalize well to the test data, where mean RMSE ratios increased to $0.943$ ($\sigma=0.134$) for azimuth and $1.003$ ($\sigma=0.150$) for elevation.

For Experiment 2, the model demonstrates more consistent performance across validation and test sets. The mean validation RMSE ratios are $0.897$ ($\sigma=0.026$) for azimuth and $0.867$ ($\sigma=0.045$) for elevation, while test data shows similar results with mean RMSE ratios of $0.953$ ($\sigma=0.061$) for azimuth and $0.943$ ($\sigma=0.088$) for elevation.

We note that models were trained with different numbers of features $k$, and these results are from the model with the best performance on the validation data ($k$ can be different for each fold).
The results show that the model's performance on the validation data is very good in both Experiment 1 and 2.
However, this does not generalize well to the test data, with the performance on the test data for Experiment 1 being significantly worse than the current model for all folds.
In contrast, for Experiment 2, the performance on the test data is better than the current model for most of the folds.

Tables \ref{tab:exp1_minvall_all} and \ref{tab:exp2_minval_all} presents the same results for the model predicting the offsets of all instruments.
This model exhibits similar trends, although its performance on the test data is not as good as the NFLASH230 model. In Experiment 1, the mean validation RMSE ratios were $0.879$ ($\sigma=0.047$) for azimuth and $0.904$ ($\sigma=0.045$) for elevation, while test performance shows mean RMSE ratios of $1.061$ ($\sigma=0.222$) for azimuth and $1.001$ ($\sigma=0.088$) for elevation. Experiment 2 shows improved consistency, with mean validation RMSE ratios of $0.929$ ($\sigma=0.026$) for azimuth and $0.888$ ($\sigma=0.062$) for elevation, and test RMSE ratios of $1.014$ ($\sigma=0.118$) for azimuth and $0.958$ ($\sigma=0.023$) for elevation.

Tables \ref{tab:exp1_nfeats_nflash230}-\ref{tab:exp2_nfeats_all} present the mean RMSE ratio for all models when using the same number of features $k$ across all folds.

\begin{table}[htp]
    \centering
    \begin{tabular}{ccccc}
    & \multicolumn{2}{c}{Azimuth} & \multicolumn{2}{c}{Elevation} \\
    \hline
    Fold & Validation & Test & Validation & Test \\
    1 &               0.858 &                0.777 &               0.856 &                0.868 \\
    2 &               0.883 &                1.013 &               0.812 &                1.205 \\
    3 &               0.799 &                1.113 &               0.841 &                1.086 \\
    4 &               0.954 &                0.801 &               0.822 &                0.836 \\
    5 &               0.879 &                1.033 &               0.749 &                1.110 \\
    6 &               0.775 &                0.922 &               0.782 &                0.917 \\
    \hline
    Mean &               0.858 &                0.943 &               0.810 &                1.003 \\
    $\sigma$ &               0.064 &                0.134 &               0.040 &                0.150 \\
    \hline
    \end{tabular}
    \caption{Experiment 1, NFLASH230.}
    \label{tab:exp1_minval_nflash230}
\end{table}

\begin{table}[htp]
    \centering
    \begin{tabular}{ccccc}
    & \multicolumn{2}{c}{Azimuth} & \multicolumn{2}{c}{Elevation} \\
    \hline
    Fold & Validation & Test & Validation & Test \\
    1 &               0.856 &                1.043 &               0.910 &                1.049 \\
    2 &               0.881 &                0.976 &               0.921 &                0.992 \\
    3 &               0.924 &                0.862 &               0.891 &                1.009 \\
    4 &               0.891 &                0.967 &               0.839 &                0.830 \\
    5 &               0.908 &                0.914 &               0.816 &                0.861 \\
    6 &               0.919 &                0.957 &               0.827 &                0.915 \\
    \hline
    Mean &               0.897 &                0.953 &               0.867 &                0.943 \\
    $\sigma$ &               0.026 &                0.061 &               0.045 &                0.088 \\
    \hline
    \end{tabular}
    \caption{Experiment 2, NFLASH230.}
    \label{tab:exp2_minval_nflash230}
\end{table}

\begin{table}[htp]
    \centering
    \begin{tabular}{ccccc}
    & \multicolumn{2}{c}{Azimuth} & \multicolumn{2}{c}{Elevation} \\
    \hline
    Fold & Validation & Test & Validation & Test \\
    1 &               0.890 &                1.478 &               0.834 &                1.084 \\
    2 &               0.907 &                0.832 &               0.944 &                0.840 \\
    3 &               0.949 &                1.006 &               0.959 &                1.021 \\
    4 &               0.820 &                0.992 &               0.877 &                1.058 \\
    5 &               0.864 &                0.958 &               0.904 &                0.966 \\
    6 &               0.841 &                1.103 &               0.909 &                1.037 \\
    \hline
    Mean &               0.879 &                1.061 &               0.904 &                1.001 \\
    $\sigma$ &               0.047 &                0.222 &               0.045 &                0.088 \\
    \hline
    \end{tabular}
    \caption{Experiment 1, all instruments.}
    \label{tab:exp1_minvall_all}
\end{table}

\begin{table}[htp]
    \centering
    \begin{tabular}{ccccc}
    & \multicolumn{2}{c}{Azimuth} & \multicolumn{2}{c}{Elevation} \\
    \hline
    Fold & Validation & Test & Validation & Test \\
1 &               0.894 &                1.244 &               0.963 &                0.981 \\
2 &               0.972 &                0.928 &               0.928 &                0.968 \\
3 &               0.915 &                1.021 &               0.939 &                0.982 \\
4 &               0.936 &                0.949 &               0.823 &                0.927 \\
5 &               0.927 &                0.945 &               0.844 &                0.938 \\
6 &               0.931 &                0.994 &               0.832 &                0.952 \\
\hline
Mean &               0.929 &                1.014 &               0.888 &                0.958 \\
$\sigma$ &               0.026 &                0.118 &               0.062 &                0.023 \\
    \hline
    \end{tabular}
    \caption{Experiment 2, all instruments.}
    \label{tab:exp2_minval_all}
\end{table}

\begin{figure}[htp]
    \centering
   \includegraphics[width=\columnwidth]{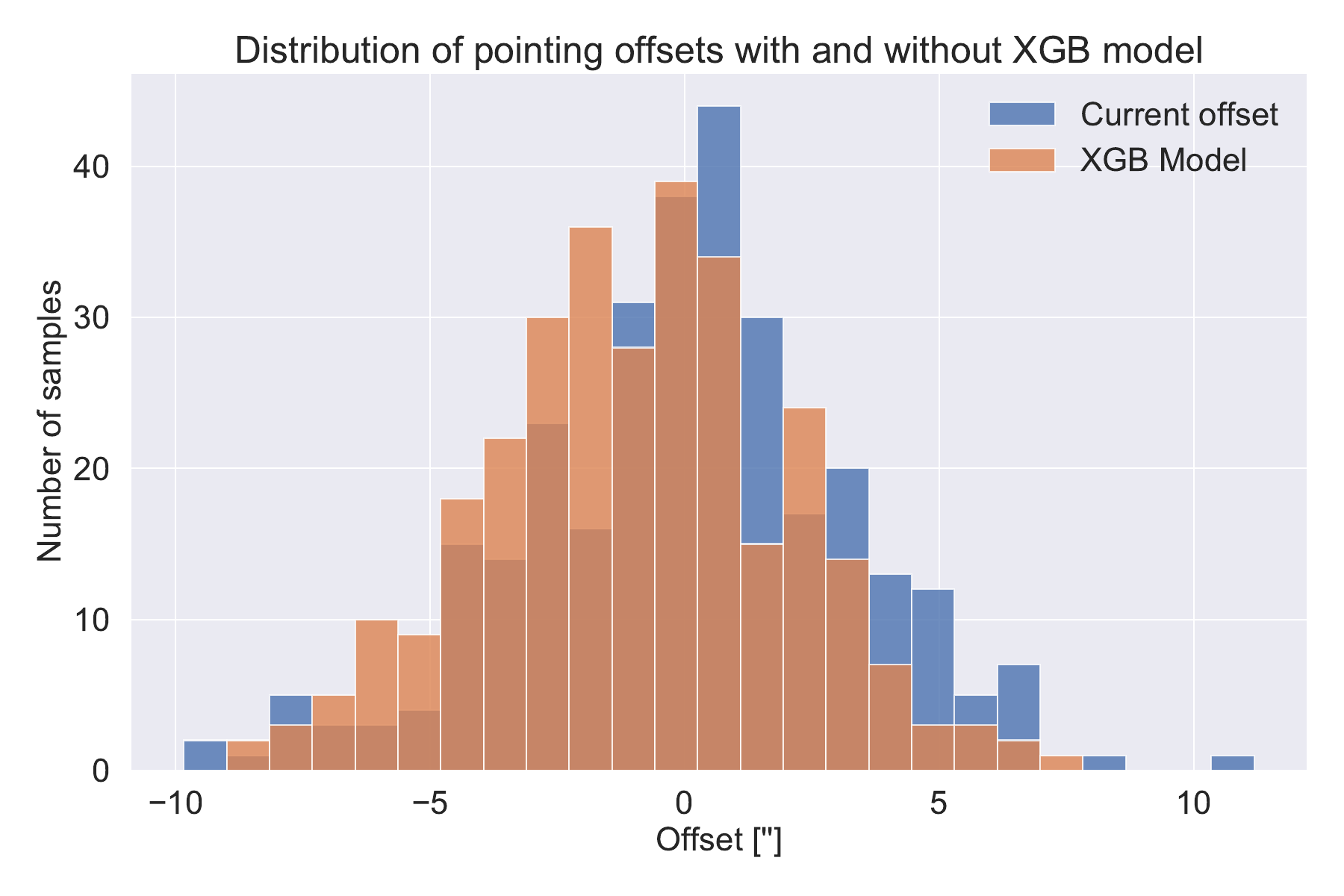}
    \caption{Distribution of NFLASH230 offsets in the azimuth coordinate, comparing without and with the machine learning model corrections. On the unseen test data, the Azimuth offsets are reduced by $4.3\%$.}
   \label{fig:hist_lastfold_nflash230_az}
\end{figure}

\begin{figure}[htp]
    \centering
   \includegraphics[width=\columnwidth]{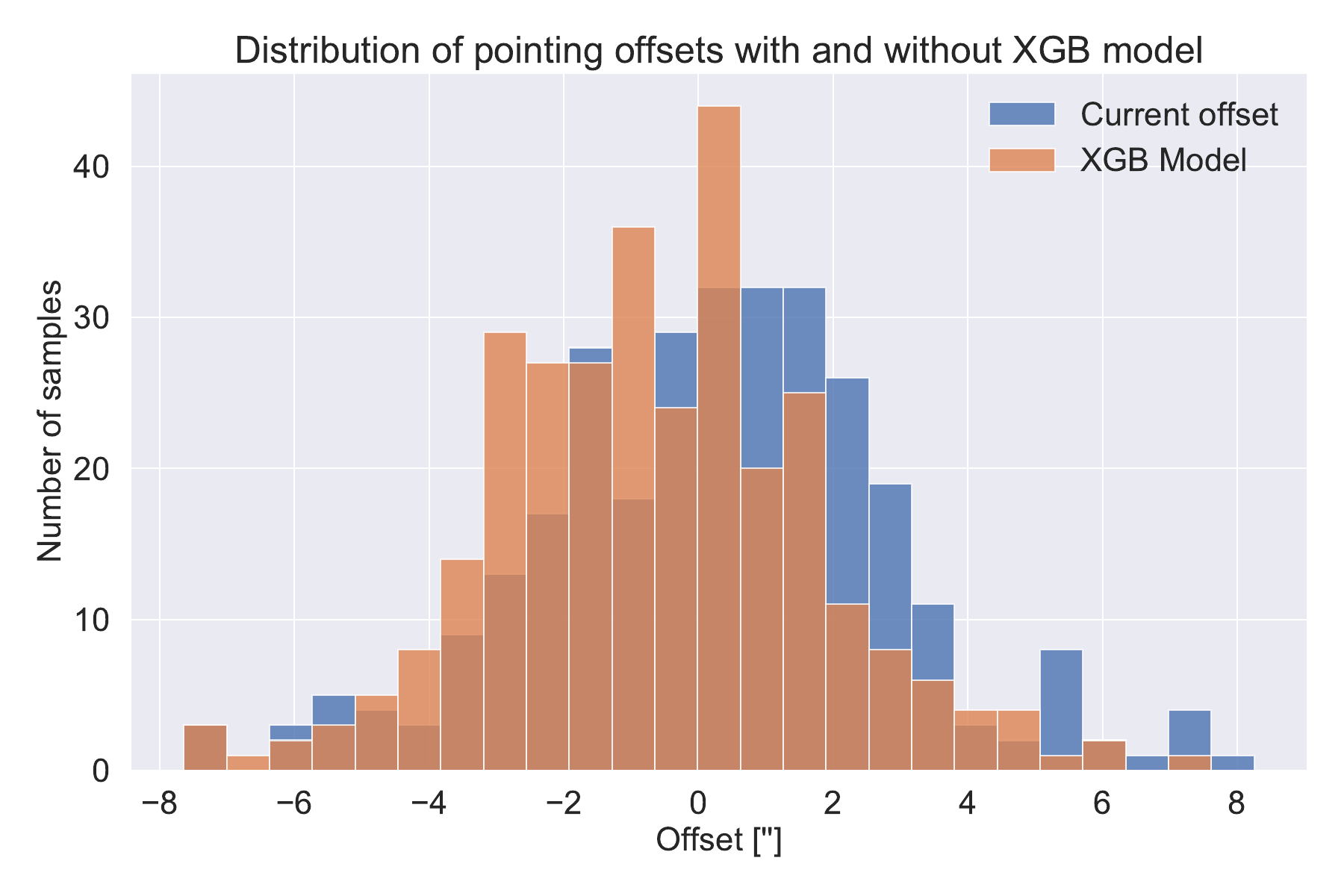}       \caption{Distribution of NFLASH230 offsets in the elevation coordinate, comparing without and with the machine learning model corrections. For the unseen test data, the elevation offsets are reduced by $9.5\%$.}
   \label{fig:hist_lastfold_nflash230_el}
\end{figure}

Figures \ref{fig:hist_lastfold_nflash230_az} and \ref{fig:hist_lastfold_nflash230_el} shows the distribution of NFLASH230 offsets in Experiment 2, with and without the machine learning model corrections.
These are the distribution of offsets of the unseen test data for the last fold. On this test set, the machine learning model reduced azimuth offsets by $4.3\%$ and elevation offsets by $9.5\%$.


\begin{table}[htp]
    \centering
    \begin{tabular}{ccccc}
        & \multicolumn{2}{c}{Azimuth} & \multicolumn{2}{c}{Elevation} \\
        k & Mean & $\sigma$ & Mean & $\sigma$ \\
        \hline
           2 &               1.017 & 0.078 &               0.978 & 0.119 \\
           5 &               1.081 & 0.225 &               0.959 & 0.082 \\
          10 &               0.989 & 0.140 &               0.938 & 0.109 \\
          20 &               0.970 & 0.159 &               0.967 & 0.109 \\
          30 &               1.005 & 0.129 &               0.989 & 0.158 \\
          40 &               0.987 & 0.182 &               0.989 & 0.146 \\
          50 &               0.976 & 0.163 &               0.987 & 0.143 \\
    \end{tabular}
    \caption{Resulting mean RMSE ratio (\ref{eq:mean_rms_compared}) on unseen test data in Experiment 1 for the model trained on NFLASH230 data, using different numbers of features $k$ in the model.}
    \label{tab:exp1_nfeats_nflash230}
\end{table}

\begin{table}[htp]
    \centering
    \begin{tabular}{ccccc}
        & \multicolumn{2}{c}{Azimuth} & \multicolumn{2}{c}{Elevation} \\
        k & Mean & $\sigma$ & Mean & $\sigma$ \\
        \hline
           2 &               1.044 & 0.132 &               1.003 & 0.059 \\
           5 &               0.998 & 0.094 &               0.995 & 0.072 \\
          10 &               1.078 & 0.197 &               0.993 & 0.089 \\
          20 &               1.083 & 0.231 &               0.978 & 0.052 \\
          30 &               1.042 & 0.222 &               0.976 & 0.065 \\
          40 &               1.010 & 0.129 &               1.006 & 0.116 \\
          50 &               1.008 & 0.132 &               0.949 & 0.053 \\
    \end{tabular}
    \caption[All instruments model results for complexity $k$ in Experiment 1]{
    Resulting mean RMSE ratio (\ref{eq:mean_rms_compared}) on unseen test data in Experiment 1 for the model trained on all data, using different numbers of features $k$ in the model.}
    \label{tab:exp1_nfeats_all}
\end{table}

\begin{table}[htp]
    \centering

    \begin{tabular}{ccccc}
        & \multicolumn{2}{c}{Azimuth} & \multicolumn{2}{c}{Elevation} \\
        k & Mean & $\sigma$ & Mean & $\sigma$ \\
        \hline
           2 &               0.941 & 0.059 &               0.955 & 0.073 \\
           5 &               0.956 & 0.077 &               0.955 & 0.083 \\
          10 &               0.966 & 0.077 &               0.955 & 0.087 \\
          20 &               0.979 & 0.140 &               0.956 & 0.080 \\
          30 &               0.965 & 0.110 &               0.937 & 0.080 \\
          40 &               0.955 & 0.081 &               0.937 & 0.075 \\
          50 &               0.956 & 0.066 &               0.934 & 0.075 \\
    \end{tabular}
    \caption{Resulting mean RMSE ratio (\ref{eq:mean_rms_compared}) on unseen test data in Experiment 2 for the model trained on NFLASH230 data, using different numbers of features $k$ in the model.}
    \label{tab:exp2_nfeats_nflash230}
\end{table}

\begin{table}[htp]
    \centering

    \begin{tabular}{ccccc}
        & \multicolumn{2}{c}{Azimuth} & \multicolumn{2}{c}{Elevation} \\
        k & Mean & $\sigma$ & Mean & $\sigma$ \\
        \hline
           2 &               0.988 & 0.069 &               0.965 & 0.014 \\
           5 &               0.986 & 0.078 &               0.960 & 0.014 \\
          10 &               1.000 & 0.112 &               0.988 & 0.069 \\
          20 &               1.006 & 0.090 &               0.973 & 0.026 \\
          30 &               1.018 & 0.118 &               0.980 & 0.042 \\
          40 &               1.019 & 0.143 &               0.950 & 0.024 \\
          50 &               1.032 & 0.178 &               0.953 & 0.032 \\
    \end{tabular}
    \caption[All instruments model results for complexity $k$ in Experiment 2]{
    Resulting mean RMSE ratio (\ref{eq:mean_rms_compared}) on unseen test data in Experiment 2 for the model trained on all data, using different numbers of features $k$ in the model.}
    \label{tab:exp2_nfeats_all}
\end{table}

\section{Discussion}
The research question addressed in this paper is whether machine learning can enhance the pointing accuracy of a radio/(sub)-mm telescope using the current pointing strategy.
To investigate this question, we explored a realistic scenario (Experiment 1) in which we trained a model on a smaller dataset corresponding to a shorter period and used it to predict the offset of consecutive scans for a following period.
We first focus on the NFLASH230 model.
The results from this experiment, presented in Table \ref{tab:exp1_minval_nflash230}, demonstrate that the model's performance on the validation data is promising,
with the RMSE ratio in the range of approximately $0.75$-$0.95$ for azimuth and elevation,
which corresponds to a $5$-$25\%$ reduction in pointing offset.
However, this performance did not transfer to the following test period,
in which the RMSE ratios range from $0.77$-$1.21$, indicating a $23\%$ reduction to a $21\%$ increase in pointing offset. Although the mean RMSE ratio is $0.943$ for azimuth and 1.00 for elevation on the test set, the performance is unreliable and inconsistent. We see similar results, though slightly worse, when predicting offsets from all instruments, for which Table \ref{tab:exp1_nfeats_all} show the result. 

There are several possible reasons for the model's poor performance on the test data.
One of the limitations of tree-based models, such as XGBoost, is that they typically do not generalize well to new data that are different from the training data,
as they predict solely based on logical conditions seen in the training data.
If the factors affecting the pointing offset change over time and the new data is different from the training data, the model will likely perform poorly. 
Another potential explanation for the poor performance could be that the dataset is too small, and the model overfits on the validation data.
The results suggest that learning the relationships in the data that affect pointing offset is challenging, and a complex model may be necessary.
In order to train a proper complex model, a larger dataset is required.
The findings also indicate that choosing the complexity of the model with the best performance on the validation data may not necessarily lead to the best performance on the test data.
We further explore this aspect by examining Tables \ref{tab:exp1_nfeats_nflash230} and \ref{tab:exp1_nfeats_all}, which lists the mean RMSE ratio on the test data when using the same number of features for all the folds, for different numbers of features. This provides an idea of the required model complexity to provide reliable models on new data, but the standard deviations are too large to make any conclusions. Forcing the number of features to be the same across folds, the performance does not become better than the existing pointing model, and we conclude that the relations to be learned are not sufficiently represented in the limited amount of data.

Moving on to Experiment 2, we test whether increasing the amount of data is beneficial despite also introducing additional complexity through slowly changing factors.
We split the data into six folds and performed cross-validation.
Since the test period was either before or between the training/validation data in time (except for the last fold), this was a less realistic experiment than Experiment $1$.
Results from this experiment indicate that a larger time period helps the model generalize better.
Table \ref{tab:exp2_minval_nflash230} show good performance on the validation data across all folds, with an $8$-$18\%$ reduced pointing offset on the validation data.
The average RMSE ratio over all folds on the test data for the NFLASH230 model in Experiment 2 is $0.953$ for azimuth and $0.943$ for elevation. 
With the standard deviations, the $95\%$ confidence intervals are $[0.833, 1.073]$ for azimuth and $[0.771,1.115]$ for elevation.
Given that the upper bound of both confidence intervals are larger than $1$ and that the data split in the experiment is not realistic in deployment, we cannot conclude that the model can reduce the pointing offset robustly and consistently. The models in Experiment 2 predicting offsets from all instruments show similar results.
We further investigate if forcing an equal number of features across all folds yields different results, shown in Tables \ref{tab:exp2_nfeats_nflash230} and \ref{tab:exp2_nfeats_all}. For the NFLASH230 model, the mean RMSE ratios consistently stay below $1$ for all model complexities. However, the standard deviations are too large to make any conclusions. When considering the all instruments model, we observe statistically significant improvements over the existing pointing model when predicting elevation offsets with model complexities $2,5,$ and $40$. For these models, the upper bound of the $95\%$ confidence interval is less than $1$.

For an unbiased result that reflects expected performance in practice, we look at the RMSE ratio of the last fold in Table \ref{tab:exp2_minval_nflash230},
showing a $4.3\%$ and $9.5\%$ reduced RMSE for azimuth and elevation, respectively.
These results are the ones visualized in the histograms in Figures \ref{fig:hist_lastfold_nflash230_az} and \ref{fig:hist_lastfold_nflash230_el}.
From the histogram showing the azimuth model results, it is not evident that the machine learning model performs better than the current model. The figure shows that the existing model is slightly biased towards positive corrections, whereas the machine learning model is slightly biased towards negative offsets.
The histogram showing the elevation offsets shows the same trends, where the machine learning model also shows an overall increase in very small offsets ($<1$~arcsec).\\

\section{Conclusion and future work}
We studied whether machine learning could enhance the pointing accuracy of a radio/(sub)-mm telescope currently applying an analytical pointing model. Our findings suggest
that, while machine learning has the potential to improve pointing accuracy, a more
extensive dataset and a complex model architecture may be necessary for consistent and robust performance. We performed two experiments, one where we trained on a smaller period and tested on unseen data in a consecutive period, and another where we trained and tested on longer periods. The results show that a longer period and more training
data yield better results than a shorter period with less training data. We also found that the NFLASH230 model provided a reduction in offset for azimuth and elevation. However, given that the experiment is not entirely realistic, we cannot conclude that the model can reduce pointing offset in a robust and consistent manner. Therefore, further research is needed to verify the performance and improvements before deploying a learned correction model in practice.

There are several potential avenues for future research
to improve the accuracy and efficiency of pointing offset prediction in radio and (sub)-mm telescopes using machine learning techniques. One key area for improvement is feature
engineering, where more informative features could be created to enhance model performance, especially when dealing with limited training data. We also suggest exploring neural networks, which offer advantages such as handling multiple outputs
and continuous fine-tuning as new data becomes available. Although our initial
tests of neural networks did not yield satisfactory results, further investigation is necessary with more extensive training data to explore the potential of neural networks. Finally, minimizing the number of
pointing scans conducted by astronomers while maintaining similar pointing accuracy
is another potential area for exploration. Overall, our study provides insights into
future research directions to optimize the performance of machine learning models for
pointing offset prediction in radio telescopes and highlights the potential for machine
learning to improve this critical aspect of radio telescope operations.

\subsection{Conflict of interest}
Authors state no conflict of interest.

\subsection{Acknowledgments}
This project has received funding from the European Union’s Horizon 2020 research and innovation program under grant agreement No 951815 (AtLAST).
This publication is based on data acquired with the Atacama Pathfinder Experiment (APEX). APEX is a collaboration between the Max-Planck-Institut fur Radioastronomie, the European Southern Observatory, and the Onsala Space Observatory.\\

We would like to thank the referees for taking the necessary time and effort to review the manuscript.

\nocite{*}
\printbibliography

\section{Appendix}

\subsection{Pointing Model Terms}
The following terms are used in the analytical pointing model.
Most of the terms described here are fitted on data collected from the optical receiver mounted in the primary mirror.
Then, the astronomers refine the terms using observations from different instruments to develop specialized pointing models for each, while most terms remain constant from the optical fit.

\subsection{Harmonic terms}
The analytical model has multiple harmonic terms, some geometrical and some empirical.
We introduce all the purely geometrical terms below, being AN, AW, CA, IA, IE, and NPAE, in addition to NRX, which is not purely geometrical but still affects pointing accuracy.
The \texttt{TPOINT} software that the APEX staff uses to develop the analytical model suggests terms that improve the model's performance on the chosen dataset.
The following terms are the empirical terms for azimuth.
\begin{align}\label{eq:analytical_az}
    \Delta Az =&  c_1 \cdot \sin{Az} + c_2 \cdot \frac{\cos{2Az}}{\cos{El}} + c_3 \cdot \cos{3Az}  \nonumber\\
    &+ c_4 \cdot \sin{2Az} + c_5 \cdot \cos{2Az} + c_6 \cdot \frac{\cos{Az}}{\cos{El}} \\
    & + c_7 \cdot \frac{\cos{5Az}}{\cos{El}}, \nonumber
\end{align}
and the terms for elevation are

\begin{align}\label{eq:analytical_el}
    \Delta El =&  c_1 \cdot \sin{El} + c_2 \cdot \cos{El}+ c_3 \cdot \cos{2Az} \nonumber\\
    &+ c_4 \cdot \sin{2Az} + c_5 \cdot \cos{3Az} + c_6 \cdot \sin{3Az} \\
    &+ c_7 \cdot \sin{4Az} + c_8 \cdot \sin{5Az} \nonumber
\end{align}

The \texttt{TPOINT} software denotes the harmonic terms in the format $Hrfci$ \cite{1994StaUN}. The list below explains the different terms.

\begin{itemize}
    \item $H$: Stands for harmonics
    \item $r$: The resulting variable, either $Az$ or $El$, denoting azimuth and elevation respectively.
    The resulting variable can also be $S$, which means the result is horizontal, or azimuth scaled by a factor $1/\cos{El}$.
    \item $f$: The harmonic function, either $S$ or $C$ denoting \textit{sine} and \textit{cosine}.
    \item $c$: The variable that the funciton $f$ is dependent on, either $Az$ or $El$.
    \item $i$: Integer value in the range $0$-$9$, denoting the frequency of the harmonic.
\end{itemize}

For example, is $\Delta Az = \text{HACA3}\cos{3Az}$ denoted as HACA3 in the \texttt{TPOINT} software.

\subsection{Az/El non-perpendicularity (NPAE)}
In an altazimuth mount, if the azimuth axis and elevation axis are not exactly at
right angles, horizontal shifts proportional to $\sin{El}$ occur. This effect is zero when pointing at the horizon and increases with elevation proportional to $1/\cos{El}$

\begin{equation}
    \Delta Az \simeq - \text{NPAE } \frac{\sin{El}}{\cos{El}}= - \text{NPAE } \tan{El},
\end{equation}
where NPAE is the horizontal displacement when pointing at Zenith.

\subsection{Left-right collimation error (CA)}
In an altazimuth mount, the collimation error is the non-perpendicularity between the nominated pointing direction and the elevation axis.
It produces a horizontal image shift given by
\begin{equation}\label{eq:pmodel_ca}
    \Delta Az \simeq -\text{CA} / \cos{El}
\end{equation}

\subsection{Azimuth and elevation index error (IA/IE)}
Index errors are the errors when pointing at origo.

The azimuth index error is 
\begin{equation}
    \Delta Az = -\text{IA},
\end{equation}

and elevation index error is
\begin{equation}\label{eq:pmodel_ie}
    \Delta El = \text{IE}
\end{equation}

\subsection{Azimuth axis misalignment (AN/AW)} 

In an altazimuth mount, misalignment of the azimuth axis north-south or east-west causes errors.
The errors caused by misalignment in the north-south are given by

\begin{equation}
    \Delta Az \simeq - \text{AN} \sin{Az} \cdot \tan{El},
\end{equation}

and

\begin{equation}
    \Delta El \simeq - \text{AN} \cos{Az},
\end{equation}
where AN is the misalignment alignment in the north-south direction.
The errors given by misalignment in east-west are given by

\begin{equation}
    \Delta Az \simeq - \text{AW} \cos{Az} \tan{El},
\end{equation}

and

\begin{equation}
    \Delta El \simeq \text{AW} \sin{Az},
\end{equation}
where AW is the misalignment alignment in the east-west direction.

\subsection{Displacement of Nasmyth rotator (NRX/NRY)}
In a Nasmyth altazimuth mount, a horizontal displacement between the elevation axis of the mount and the rotation axis of the Nasmyth instrument-rotator produces
and image shift on the sky with a horizontal component
\begin{equation}
    \Delta Az \simeq - \text{NRX},
\end{equation}
and an elevation component
\begin{equation}
    \Delta El \simeq - \text{NRX} \sin{El},
\end{equation}
where NRX is the horizontal displacement.

A vertical displacement produces an image shift on the sky with a horizontal component 
\begin{equation}
    \Delta Az \simeq - \text{NRY} \tan{El},
\end{equation}
and an elevation component
\begin{equation}
    \Delta El \simeq \text{NRY} \cos{El},
\end{equation}

In the case of APEX, $\text{NRY}=0$.

\end{document}